\documentclass{aa}
\usepackage{graphicx}
\usepackage{natbib}
\bibpunct{(}{)}{;}{a}{}{,} 

\usepackage[varg]{txfonts}
\usepackage{color}

\begin{document}

\title{A new approach to feature-based asteroid taxonomy in 3D color space: 1. SDSS photometric system}

\author{Dong-Goo Roh\inst{1} \and Hong-Kyu Moon\inst{1} \and Min-Su Shin\inst{1} \and Francesca E. DeMeo\inst{2}}

\institute{
Korea Astronomy and Space Science Institute, Dajeon 34055, Republic of Korea \\ \email{rrdong9@kasi.re.kr}\label{inst1} \and Department of Earth, Atmospheric, and Planetary Sciences, Massachusetts Institute of Technology, 77 Massachusetts Avenue, Cambridge, MA 02139 USA 
}

\abstract{The taxonomic classification of asteroids has been mostly based on 
spectroscopic observations with wavelengths spanning from the visible (VIS) 
to the near-infrared (NIR). 
VIS-NIR spectra of  $\sim$2500 asteroids have been obtained since the 1970s;  the Sloan Digital Sky Survey (SDSS)  Moving Object Catalog 4 (MOC 4) 
was released with  $\sim$4 $\times$ 10$^{5}$ measurements of asteroid positions 
and colors in the early 2000s. 
A number of works then devised methods to classify these data within the framework of existing taxonomic systems.
Some of these works,  however, used 2D parameter space 
(e.g., gri slope vs. z-i color) that displayed a continuous distribution of 
clouds of data points resulting in boundaries that were  artificially defined.
We introduce here a more advanced method to classify asteroids based on existing systems.
This approach is simply represented by a triplet of SDSS colors. 
The distributions and memberships of each taxonomic type are determined by 
machine learning methods in the form of both unsupervised and semi-supervised learning. 
We apply our scheme to MOC 4 calibrated with VIS-NIR reflectance spectra. We successfully separate seven different taxonomy 
classifications (C, D, K, L, S, V, and X) with which 
we have a sufficient number of spectroscopic datasets.
We found the overlapping regions of taxonomic types in a 2D plane were separated with relatively clear boundaries in the 3D space newly defined in this work.
Our scheme explicitly discriminates between different taxonomic types (e.g., K and X types), which is an improvement over existing systems.
This new method for taxonomic classification has a great deal of 
scalability for asteroid research, such as space weathering in the S-complex, 
and the origin and evolution of asteroid families. We present the structure of the asteroid belt, and describe the orbital distribution 
based on our newly assigned taxonomic classifications. It is also possible 
to extend the methods presented here to other photometric systems, such as the Johnson-Cousins and LSST filter systems.}

\keywords{minor planets, asteroids: general --- techniques: photometric --- methods: statistical}

\maketitle

\section{Introduction}

Taxonomy is defined as the practice and science of classification; the word is derived from the Greek roots taxis (order or arrangement) and nomos (law or science). 
In a broad sense it refers to the classification of things or concepts, and the principles underlying a system of classification. 
A system usually has a hierarchical structure with subtypes or subclasses.  
The expression naturally applies to the classification of asteroids as they exhibit a variety of spectral properties linked to their orbits. 
In the mid-1900s  \citet{Kitamura59} discovered a color gradient among asteroids, in the sense that distant objects (> 3.0 AU) were systematically bluer than those found closer to the Sun (< 2.3 AU). 
This was later confirmed by \citet{Chapman71} based on Johnson UBV photometry of dozens of asteroids. 
His work is considered the very beginning of asteroid taxonomy \citep{Tedesco89}.  
The studies followed by \citet{Zellner73} revealed that asteroids fall into at least two principal classes with seemingly distinct physical properties. 
They introduced C and S nomenclature to characterize the surface properties of asteroids which we still use today. 
It was later confirmed and strengthened by \citet{Chapman75}.

 The list of known principal classes has been gradually expanded to include more than a dozen  different major and minor asteroid types. 
The number of these types (singular taxon; plural taxa) has continued to grow as more observational data become available and an old scheme is replaced with a more sophisticated one. 
In 1984 \citet{Tholen84} developed an extended and powerful taxonomic system to further classify 14 taxa based on the Eight-Color Asteroid Survey (ECAS) \citep{Zellner85} 
which includes a photometric dataset of 589 asteroids.
 
 In the early 2000s the number of asteroids with taxon reached $\sim$2000 thanks to dedicated spectroscopic surveys of asteroids  \citep{Bus02b, Mothe03}.
 \citet{Bus02b} and  \citet{Lazzaro04} independently conducted well-designed large-scale surveys and measured visible spectra for 1447 and 820 asteroids, respectively. 
The former is called Phase II of the Small Main-belt Asteroid Spectroscopic Survey (SMASSII) and the latter the Small Solar System Objects Spectroscopic Survey (S$^{3}$OS$^{2}$). 
\citet{Bus02a} established an extended classification system maintaining the frameworks of the existing taxonomies. 
They defined three major groups called the C-, S-, and X-complexes, that preserve the classical definitions of the above-mentioned asteroid groups. 
They finally adopted a total of 26 taxa depending on the presence, absence, or degree of certain spectral features (e.g., spectral slope shortward of 0.75 microns and the absorption band depth around 1 micron) \citep{Bus02a}. 
More recently, the Bus-DeMeo classification system \citep{DeMeo09} used $\sim$ 400 visible (VIS) and near-infrared (NIR) spectra to extend the Bus-Binzel taxonomy system to NIR wavelengths.
For decades asteroid taxonomy has been used to characterize an asteroid's individual physical properties. 
Combining taxonomy with large-scale surveys has unlocked the door to study the makeup, origins, and evolution of the whole population \citep{Ivezic01,Mainzer11,Masiero11,Popescu18}.

Many studies then used the framework of existing taxonomies to classify the sample of $\sim$ 4$\times$10$^{5}$ asteroids listed in the Sloan Digital Sky Survey (SDSS)  Moving Object Catalog 4 (MOC 4) \citep{Ivezic01} dataset.
In their innovative studies \citet{Ivezic02} and \citet{Parker08} introduced a* vs. i-z color space to distinguish colors assigned to C, S, and V taxonomic types
to further investigate the nature of asteroid dynamical families in proper orbital element space.
Using SDSS MOC4 data, \citet{Carvano10} suggested a new taxonomy
that is compatible with previous classification schemes, 
while \citet{Hasselmann15} redefined a different taxonomy
independent from the preceding ones.
\citet{DeMeo13} made use of $\sim$ 400   VIS-NIR
 spectra as control points to apply their taxonomic scheme to SDSS data.
However, it turned out that the boundaries of major complexes and subclasses are  artificially defined. 
This is due to the fact that the ensemble of asteroid spectra of hundreds of thousands asteroids in  two dimensional (2D) parameter space (e.g., slope vs. depth or gri slope vs. z-i color) used in that work appears to be a continuous distribution of clouds of data points without any clear boundaries.

In this work we seek to improve asteroid classification based on multi-band photometry
such as SDSS by defining mathematically discrete boundaries
for the existing taxonomic types by introducing a third dimension for classification and by applying clustering techniques. In Section 2 we describe our dataset and a third dimension we define to improve classification. In Section 3 we present our clustering methods and our taxonomic results. In Section 4 we describe the improvements in classification seen with our additional third dimension and present the distribution of asteroids in the main belt based on our classification results.

\section{The concept of the 3D taxonomy}
\subsection{Dataset}

\begin{figure}
  \centering
  \includegraphics[width=\hsize]{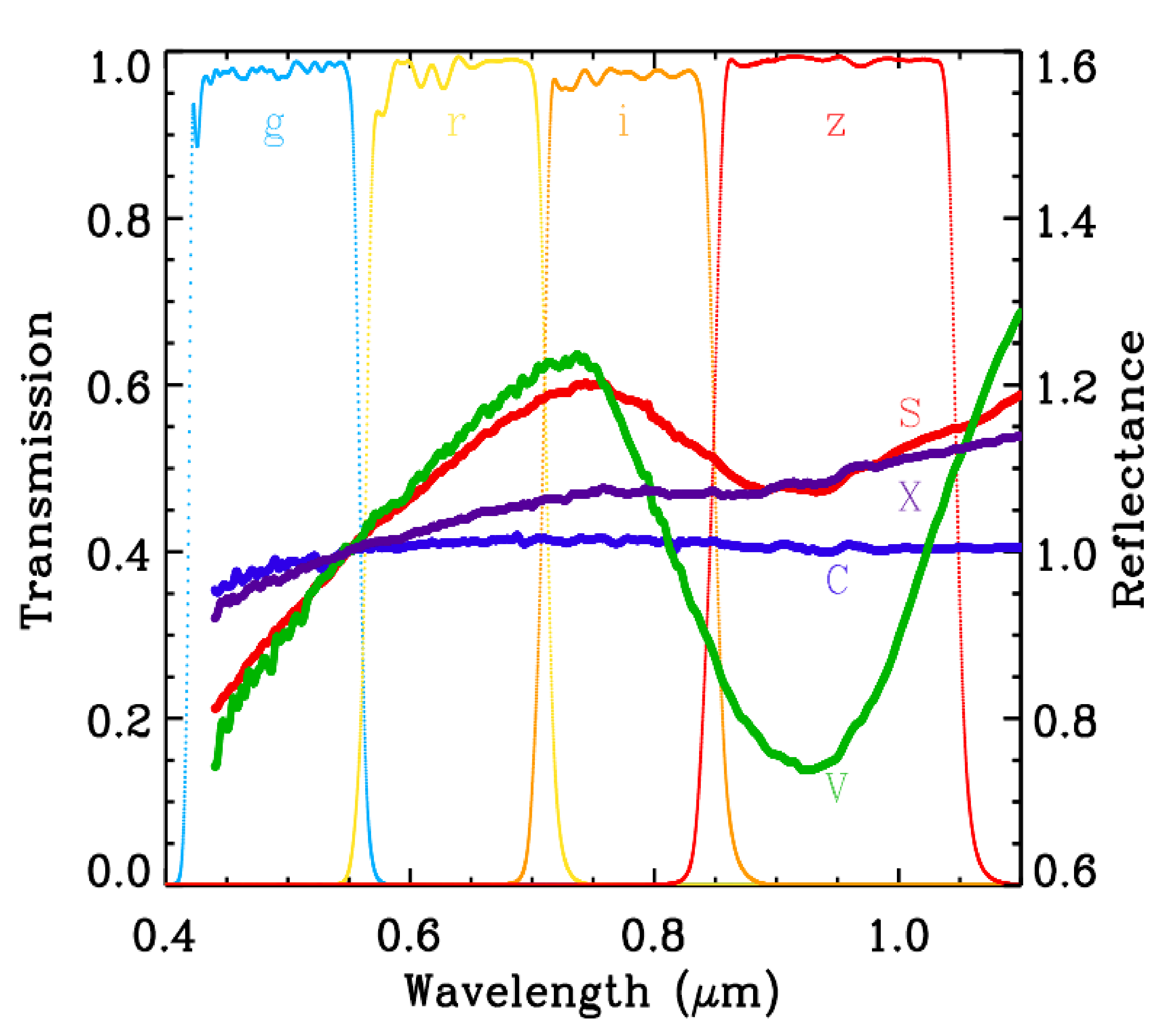}
  \caption{SDSS filter transmission curve and average spectra of the Bus-DeMeo representative types. The spectra are normalized at 550 nm.}
  \label{fig:SDSS_filter_tax}
\end{figure}

\begin{figure}
  \centering
  \includegraphics[width=\hsize]{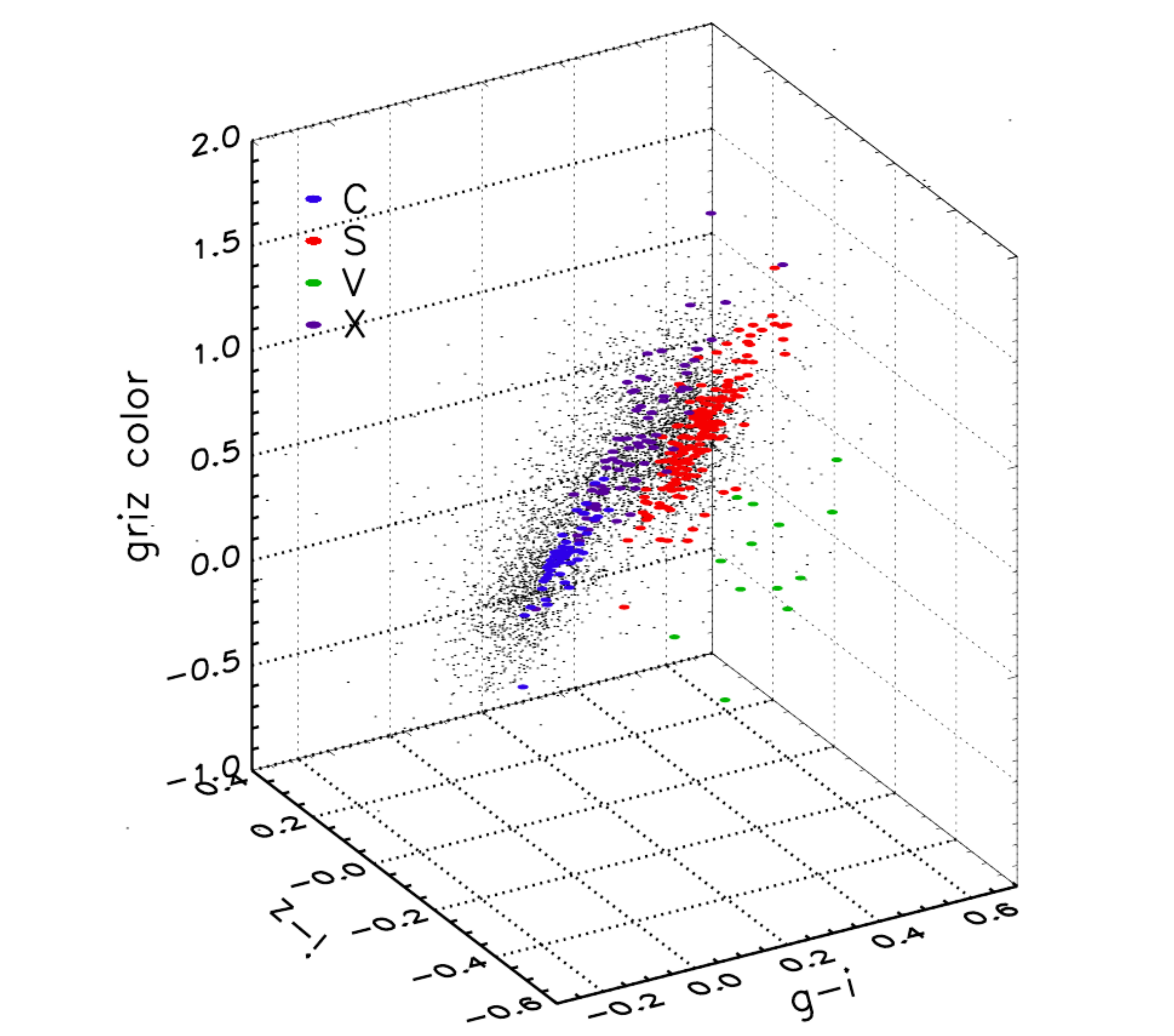}
  \caption{Three-dimensional convolution color diagram of the \citet{DeMeo13} spectra and MOC4 dataset. Black background dots represent MOC4 photometric data without their spectral measurements.}
  \label{fig:3D_color_space}
\end{figure}

We based our new asteroid taxonomy on the spectra of 318 asteroids using the Bus-DeMeo classification method \citep{DeMeo09}.
Of the 371 they used, we exclude  the spectra that did not sufficiently
cover the g band.
The Bus-DeMeo classification basically follows the  \citet{Bus02b} methodology,
except for a difference in the number of subclasses because the Bus-DeMeo taxonomy was defined using a larger range of wavelengths (VIS and NIR).
The set of reflectance spectra used to define the Bus-DeMeo classification constitutes
reference values to set the boundary conditions of subclasses in
this study. 
The number of spectra used for each type is as follows:
5 for A, 3 for B, 44 for C, 13 for D, 12 for K, 15 for L, 4 for Q, 1 for R, 173 for S, 4 for T, 13 for V, and 31 for X.

The SDSS has proved to be useful in planetary science by providing photometric measurements of the significantly increased number of small  Solar System bodies. MOC4 includes over 470000 moving objects in the Solar System \citep{Ivezic02}. 
However, some of the data have large uncertainties in the photometric measurements. 
Our sample is thus selected with the following criteria. 
First, we exclude SDSS u-band data because the  DeMeo spectral data does not cover the corresponding wavelengths. 
Then we select either numbered objects or objects with provisional designations to restrict the sample to those with higher precision of their proper orbital elements. 
Following DeMeo's work, we choose observations that are sufficiently bright to be considered reliable, g < 22.2, r < 22.2, i < 21.3, and z < 20.5, which are the limiting magnitudes for 95$\%$ completeness \citep{Ivezic01}. 
In addition, data with photometric uncertainty smaller than 0.05 are included except for the u filter. 
To avoid possible contamination by a crowded stellar field in the Milky Way, 
we include only objects greater than 15 degrees in galactic latitude, which is a more stringent condition than used in \citet{DeMeo13} and other previous works.
This final constraint excludes a significant fraction ($\sim$75 \%) of asteroids in the dynamical plane, especially for the ones with low orbital inclinations.
Our final dataset is a sample of 4213 asteroids from SDSS MOC4, which are regarded as being free of significant photometric errors.

\subsection{Transformation from 2D reflectance to 3D colors}
Traditionally for asteroid taxonomy, Principal Component Analysis has been the primary technique used to distinguish classes. 
\citet{Bus02a} used a statistical method to determine taxonomic classes based on the shape of the reflectance spectra, namely 
the spectral slope shortward of 0.75 um and the depth of 1 um absorption band, which are the principal components. 
\citet{DeMeo09} used the principal components (slope and band depth) obtained with
VIS-NIR spectra of 371 asteroids. \citet{DeMeo13} convert SDSS colors to their spectral analogs in the 2D color
space (slope and z-i color) to classify over 34000 unique objects
using the color space information of those 371 asteroids.
Similarly, in this work we convert reflectance spectra of \citet{DeMeo09} sample to SDSS colors by convolving them with the SDSS filter transmission curves (see Figure \ref{fig:SDSS_filter_tax}).
We compute the convolved flux ratio in the g band and i band to have ($g-i$) color, while we compute the convolved flux ratio in i and z band, to obtain ($i-z$) color which are analogs of principal components defined by \citet{Bus02b} and \citet{DeMeo09}. 

In order to overcome the limitation set by the previous studies (artificially drawn boundaries), we introduce the griz color, a new color for asteroid taxonomy; the sum of $g-r$, $g-i$, and $g-z$ colors represents the flux values of the normalized reflectance in the SDSS bands. 
We then classify 4213 MOC4 objects based on the distribution of those in the 3D 
color space defined with the three colors  shown in Figure \ref{fig:3D_color_space}. 
This newly defined color adds an extra dimension. The mathematical meaning and benefits of applying the griz color is further described in Section 4.1.

\section{Clustering for taxonomy assignment}

\begin{figure}
  \centering
  \includegraphics[width=\hsize]{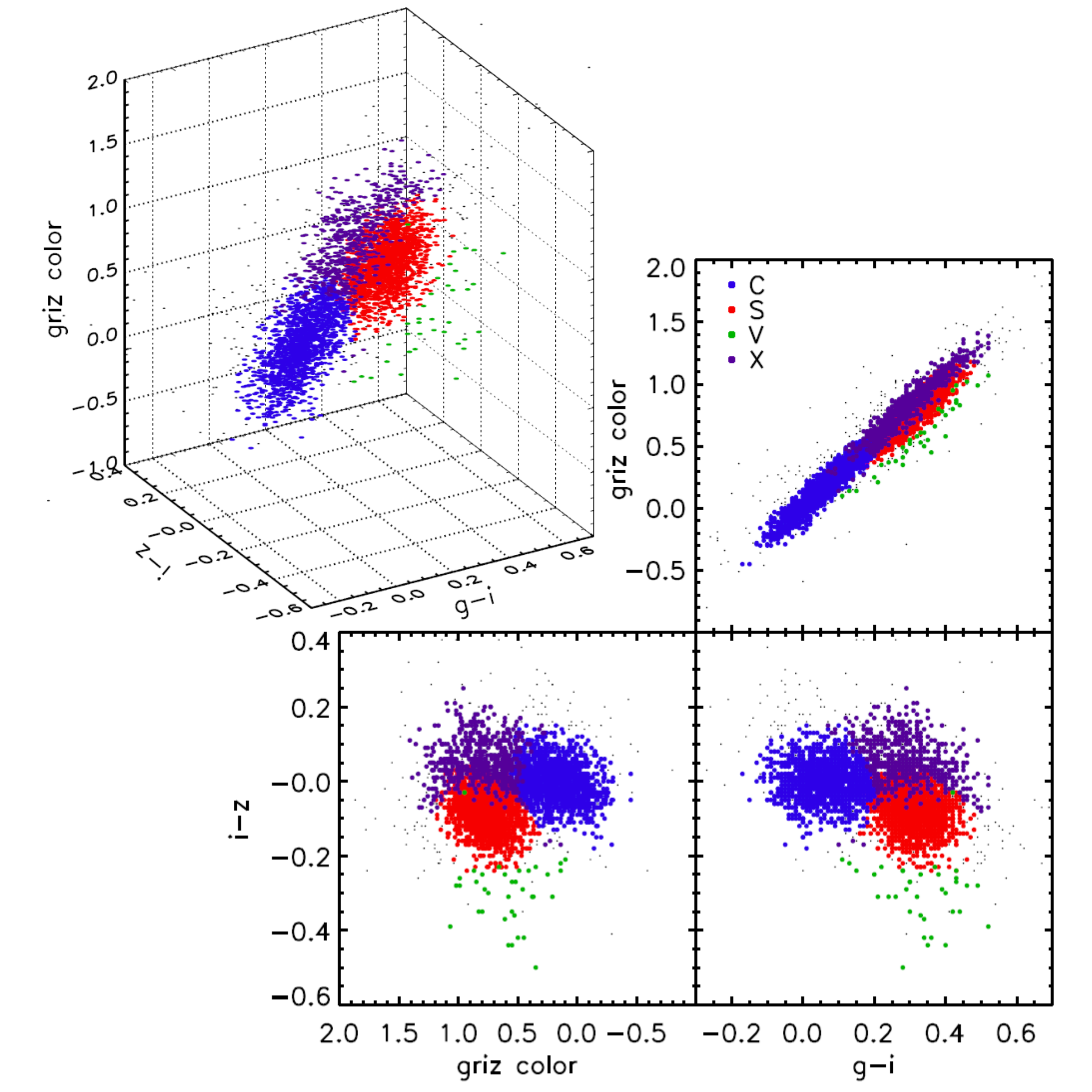}
  \caption{Three-dimensional clustering diagram for MOC4 without any information of known asteroid types and colors in the method A1 taxonomy assignment with the posterior dissimilarity matrix. 
Different colors correspond to different taxonomy classification memberships: 
C type (blue), S type (red), V type (lime green), X type (purple), and Unassigned (small black).}
  \label{fig:methodA_membership}
\end{figure}

\begin{figure}
  \centering
  \includegraphics[width=\hsize]{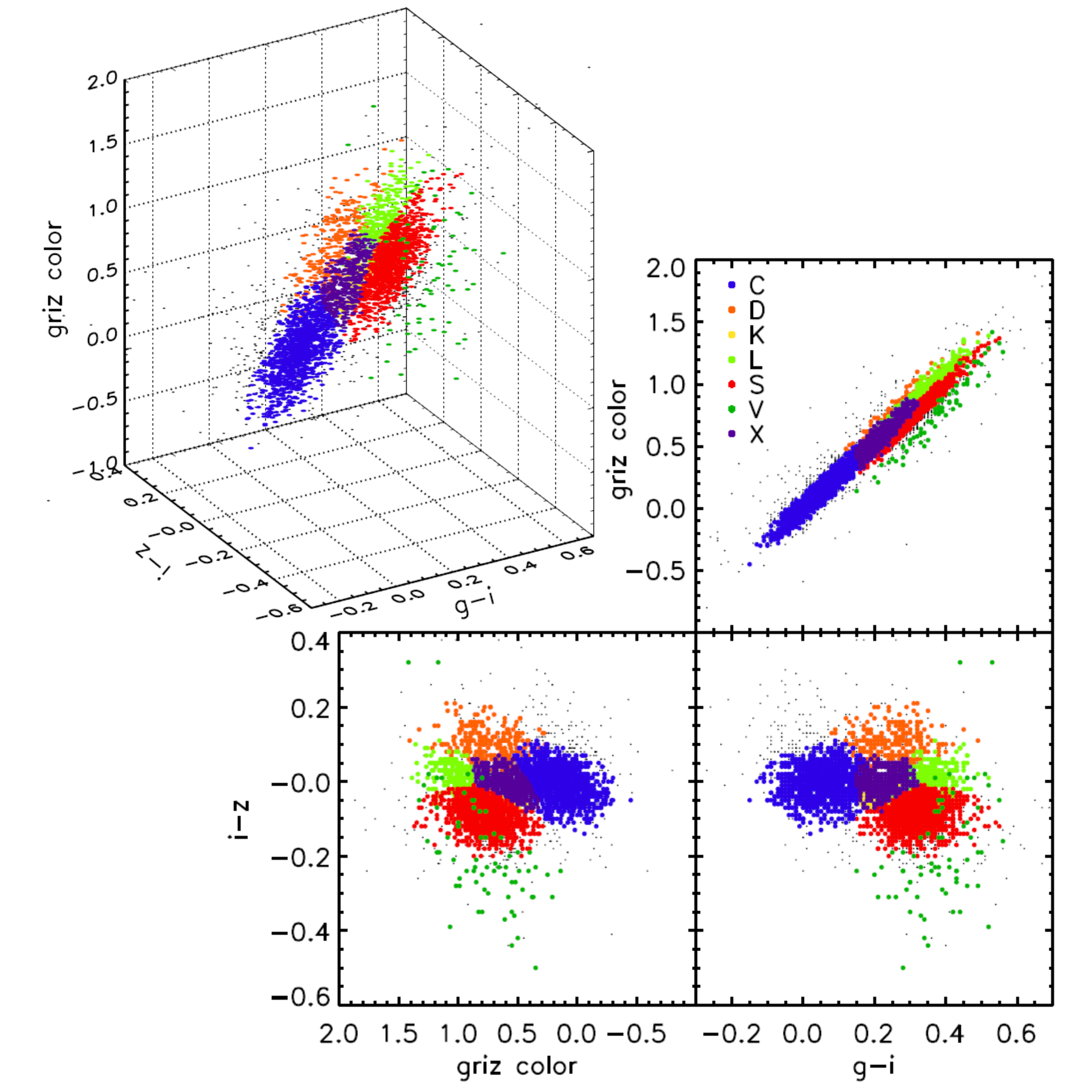}
  \caption{Three-dimensional clustering diagram for MOC4 with known asteroid types and colors of DeMeo dataset in the method B taxonomy assignment. Different colors correspond to different taxonomy classification memberships:  C (blue), D (orange red), K (yellow), L (lime), S (red), V (lime green), X (purple), and Unassigned (small black).}
  \label{fig:methodB_membership}
\end{figure}

Clustering is one kind of machine learning method to identify 
structures in a given dataset;  there are numerous methods of clustering. 
The exploration of the data in the 3D color space as shown in 
Figure \ref{fig:3D_color_space} informs us that the 
structure traced by the objects with the known taxonomy types can be well 
described by Gaussian shapes in the color space. As presented below, 
we chose Gaussian mixture models as an adequate clustering model to deduce 
taxonomy types. The interpretation of the inferred mixture results is not 
difficult and is easily understandable because of the model's concise and robust 
prescription and implementation.

We applied two clustering methods using Gaussian mixtures 
to identify known taxonomy types 
in the 3D color space. 
The two methods have been successfully used in astronomy 
\citep{Shin09, Shin12, Shin18}. 
The first method (hereafter method A) uses an infinite Gaussian mixture 
model to  describe  the distribution of objects as a mixture of 
multiple Gaussian distributions in  multi-dimensional color space 
without fixing the number of the components beforehand. 
This method simply tries to find a concentration of data in  
multi-dimensional color space even though  we already know the taxonomy 
types of some objects with measured colors. 
Therefore, if there are 
not enough data to be identified as concentrated clusters,  method A 
fails to recover these low-density clusters.
The second method (hereafter method B) 
is a finite Gaussian mixture model that needs to predefine 
the number of Gaussian mixture components, and we adopt  method B 
as a semi-supervised machine learning method \citep{SSLbook} 
that uses the colors of a few objects 
with known taxonomy types as a guide to infer the mixture properties. 
 In  method B new types cannot be found since 
the number of clusters is fixed as the number of the given objects
with the known taxonomy types.
After estimating the mixture of Gaussian components that describe 
the color distribution of the input data, 
the two methods assign taxonomy types (i.e., cluster 
memberships) differently, as described later. 
If some objects have consistent taxonomy assignments 
between these multiple methods, we can consider that their derived 
taxonomy types are more reliable than others.

Our usage of the unsupervised and semi-supervised learning methods requires 
our interpretation of the clustering results in deriving 
the identification of the clusters in terms of the taxonomy types and 
evaluating the clustering quality with the comparison 
to the distribution of the known taxonomy types in the color space  \citep[e.g.,][]{Kiar17}.

\subsection{Methods}

The color distribution of the input data are described as multiple 
3D Gaussian mixtures by the following equation:

\begin{equation}\label{eq:GMM}
  \begin{split}
    p({\bf x}) & 
    = ~ \sum_{k=1}^{K} w_{k}\frac{1}{(2\pi)^{D / 2}\vert\Sigma_{k}\vert^{1/2}} 
\exp\left[-\frac{1}{2}({\bf x} 
- {\bf \mu_{k}})^{T}\Sigma_{k}^{-1}({\bf x} - {\bf \mu_{k}})\right] \\
 & = ~ \sum_{k=1}^{K} w_{k} N({\bf x} | {\bf \mu_{k}, \Sigma_{k}}).
  \end{split}
\end{equation}

\noindent Here $k$ is an index over $K$ mixtures, ${\bf x}$ 
is the vector of object colors,
$D$ is the dimension of the color space (i.e., $D ~ = ~ 3$ in this paper),
and $w_{k}$ is a mixing fraction. 
The centers and covariances of the mixture components 
are $\mu_{k}$ and $\Sigma_{k}$, respectively.
Method A assumes that the number of components is infinity 
by considering  $K$ as a stochastic parameter that needs to be inferred, 
while  method B fixes $K$ to a specific value. 
In our case,  $K$ is equal to 12 in  method B 
since our data consists of 12 taxonomic types and 
includes known samples of the 12 types.

\begin{table}
\caption{Taxonomy types in  method A with the dissimilarity matrix}
\label{tab:methodA_raw_membership}
\centering
\begin{tabular}{c c}
\hline
\hline
Object name & Taxonomy type\\
\hline
1999 NG53 & S \\
Kenzo & S \\
1999 NH54 & S \\
2005 SA221 & X \\
1999 RP116 & X \\
3090 P-L & X \\
1999 NO55 & S \\
1999 XV242 & S \\
Marlu & C \\
Sansyu-Asuke & S \\
1996 VO4 & C \\
1999 VJ14 & S \\
Priestley & X \\
Halaesus & X \\
Belinskij & C \\
2000 HC36 & S \\
2000 HA41 & X \\
1992 SU & C \\
Neuvo & S \\
2000 GT136 & S \\
\hline
\end{tabular}
\tablefoot{This table is published in its entirety in  machine-readable format.
A portion is shown here for guidance regarding its form and content.}
\end{table}

\begin{table*}
\caption{MAP parameter estimation of mixture components 
in  method A\label{tab:methodA_MAP}}
\centering
\begin{tabular}{cccc}
\hline
\hline
Cluster number (Taxonomy type)\tablefootmark{a} & $w$ & $\mu$ & $\Sigma\tablefootmark{b}$ \\
\hline
1 (C) & 3.85E-01 & (7.29E-02, 5.68E-03, 2.05E-01) & 
$ \begin{pmatrix} 
6.35E-03 & 1.41E-04 & 1.67E-02 \\ 
\cdots & 2.86E-03 & 3.38E-03 \\ 
\cdots & \cdots & 4.84E-02 \\ 
\end{pmatrix} $\\
2 (S) & 4.74E-01 & (3.05E-01, -6.75E-02, 7.57E-01) & 
$ \begin{pmatrix} 
4.07E-03 & -7.34E-04 & 9.97E-03 \\ 
\cdots & 3.50E-03 & 1.85E-03 \\ 
\cdots & \cdots & 2.97E-02 \\ 
\end{pmatrix} $\\
3 (-) & 6.94E-02 & (2.00E-01, -2.19E-03, 5.38E-01) & 
$ \begin{pmatrix} 
5.24E-02 & -2.64E-02 & 1.05E-01 \\ 
\cdots & 2.73E-02 & -3.41E-02 \\ 
\cdots & \cdots & 2.52E-01 \\ 
\end{pmatrix} $\\
4 (X) & 6.36E-02 & (3.20E-01, 7.55E-02, 9.16E-01) & 
$ \begin{pmatrix} 
6.94E-03 & -2.59E-03 & 1.56E-02 \\ 
\cdots & 3.87E-03 & -3.09E-03 \\ 
\cdots & \cdots & 3.98E-02 \\ 
\end{pmatrix} $\\
5 (-) & 4.17E-03 & (5.34E-01, 1.15E-01, 1.55E+00) & 
$ \begin{pmatrix} 
3.82E-01 & -2.40E-01 & 7.97E-01 \\ 
\cdots & 1.93E-01 & -4.12E-01 \\ 
\cdots & \cdots & 1.90E+00 \\ 
\end{pmatrix} $\\
6 (V) & 3.74E-03 & (3.31E-01, -3.15E-01, 6.07E-01) & 
$ \begin{pmatrix} 
4.31E-03 & -1.97E-04 & 1.11E-02 \\ 
\cdots & 5.29E-03 & 4.03E-03 \\ 
\cdots & \cdots & 3.57E-02 \\ 
\end{pmatrix} $\\
7 (-) & 2.71E-05 & (-1.28E+00, 1.61E+00, 4.84E+00) & 
$ \begin{pmatrix} 
2.61E-02 & -1.54E-02 & 5.37E-02 \\ 
\cdots & 1.11E-02 & -3.01E-02 \\ 
\cdots & \cdots & 1.15E-01 \\ 
\end{pmatrix} $\\
8 (-) & 1.50E-04 & (1.37E+00, 2.73E-01, 2.50E-01) & 
$ \begin{pmatrix} 
2.56E-02 & -9.97E-03 & 5.27E-02 \\ 
\cdots & 1.27E-02 & -1.52E-02 \\ 
\cdots & \cdots & 1.15E-01 \\ 
\end{pmatrix} $\\9 (-) & 1.24E-04 & (9.85E-01, 2.62E+00, -2.52E+00) & 
$ \begin{pmatrix} 
3.19E-02 & -8.50E-03 & 6.19E-02 \\ 
\cdots & 5.87E-03 & -1.46E-02 \\ 
\cdots & \cdots & 1.26E-01 \\ 
\end{pmatrix} $\\10 (-) & 1.53E-04 & (3.81E-01, 1.63E+00, -1.09E+00) & 
$ \begin{pmatrix} 
1.53E-02 & -7.24E-03 & 3.00E-02 \\ 
\cdots & 6.50E-03 & -9.81E-03 \\ 
\cdots & \cdots & 6.77E-02 \\ 
\end{pmatrix} $\\11 (-) & 4.22E-05 & (1.02E+00, -1.85E+00, 1.96E+00) & 
$ \begin{pmatrix} 
2.31E-02 & 7.29E-03 & 5.08E-02 \\ 
\cdots & 1.94E-02 & 1.70E-02 \\ 
\cdots & \cdots & 1.18E-01 \\ 
\end{pmatrix} $\\12 (-) & 3.52E-05 & (2.47E-01, -6.99E-02, -2.77E+00) & 
$ \begin{pmatrix} 
1.18E-02 & -7.83E-03 & 2.70E-02 \\ 
\cdots & 1.06E-02 & -1.48E-02 \\ 
\cdots & \cdots & 6.50E-02 \\ 
\end{pmatrix} $\\13 (-) & 1.50E-04 & (6.70E-01, -5.62E-01, -3.06E+00) & 
$ \begin{pmatrix} 
8.88E-03 & 3.88E-03 & 3.37E-02 \\ 
\cdots & 1.19E-02 & 3.05E-02 \\ 
\cdots & \cdots & 1.56E-01 \\ 
\end{pmatrix} $\\
\hline
\end{tabular}
\tablefoot{
\tablefoottext{a}{Clusters without corresponding taxonomy types are 
shown as -.}
\tablefoottext{b}{Because $\Sigma$ is a symmetric matrix, we do not present 
a lower triangular part of matrix elements in this table.}}
\end{table*}

\begin{table*}
\caption{Taxonomy types in  method A with the MAP estimation of the parameters 
\label{tab:methodA_MAP_membership}}
\centering
\begin{tabular}{ccccccc}
\hline
\hline
Object name &
Cluster number (Taxonomy type)\tablefootmark{a} & 
  \multicolumn{5}{c}{Membership weights\tablefootmark{b}} \\
& & (1) & 
(2) & $\cdots$ &
(12) & (13) \\
\hline
1999 NG53 & 2 (S) & 1.0E-02 & 9.8E-01 & $\cdots$ & 0.0E+00 & 0.0E+00 \\
Kenzo & 2 (S) & 3.2E-03 & 8.1E-01 & $\cdots$ & 0.0E+00 & 0.0E+00 \\
1999 NH54 & 2 (S) & 7.9E-05 & 9.6E-01 & $\cdots$ & 0.0E+00 & 0.0E+00 \\
2005 SA221 & 4 (X) & 1.5E-03 & 6.0E-02 & $\cdots$ & 0.0E+00 & 0.0E+00 \\
1999 RP116 & 4 (X) & 2.2E-04 & 1.2E-05 & $\cdots$ & 0.0E+00 & 0.0E+00 \\
3090 P-L & 1 (C) & 4.6E-01 & 2.3E-02 & $\cdots$ & 0.0E+00 & 0.0E+00 \\
1999 NO55 & 2 (S) & 3.5E-03 & 9.9E-01 & $\cdots$ & 0.0E+00 & 0.0E+00 \\
1999 XV242 & 2 (S) & 6.0E-04 & 9.8E-01 & $\cdots$ & 0.0E+00 & 0.0E+00 \\
Marlu & 1 (C) & 9.9E-01 & 7.8E-04 & $\cdots$ & 0.0E+00 & 0.0E+00 \\
Sansyu-Asuke & 2 (S) & 1.3E-03 & 9.9E-01 & $\cdots$ & 0.0E+00 & 0.0E+00 \\
1996 VO4 & 1 (C) & 7.8E-01 & 1.5E-01 & $\cdots$ & 0.0E+00 & 0.0E+00 \\
1999 VJ14 & 2 (S) & 2.9E-05 & 8.8E-01 & $\cdots$ & 0.0E+00 & 0.0E+00 \\
Priestley & 2 (S) & 1.5E-05 & 7.4E-01 & $\cdots$ & 0.0E+00 & 0.0E+00 \\
Halaesus & 4 (X) & 7.4E-02 & 6.7E-02 & $\cdots$ & 0.0E+00 & 0.0E+00 \\
Belinskij & 1 (C) & 9.1E-01 & 3.9E-02 & $\cdots$ & 0.0E+00 & 0.0E+00 \\
2000 HC36 & 2 (S) & 1.3E-03 & 9.8E-01 & $\cdots$ & 0.0E+00 & 0.0E+00 \\
2000 HA41 & 2 (S) & 5.4E-02 & 5.6E-01 & $\cdots$ & 0.0E+00 & 0.0E+00 \\
1992 SU & 1 (C) & 9.8E-01 & 4.8E-03 & $\cdots$ & 0.0E+00 & 0.0E+00 \\
Neuvo & 2 (S) & 2.1E-04 & 9.7E-01 & $\cdots$ & 0.0E+00 & 0.0E+00 \\
2000 GT136 & 2 (S) & 1.2E-04 & 9.3E-01 & $\cdots$ & 0.0E+00 & 0.0E+00 \\
\hline
\end{tabular}
\tablefoot{This table is published in its entirety in  machine-readable format. 
A portion is shown here for guidance regarding its form and content.
\tablefoottext{a}{Types except for C, S, V, and X types 
are not shown here  since their assignments are not reliable in the clustering results.}
\tablefoottext{b}{The membership weights are defined by 
  $\hat{w}_{k}  N({\bf x} | {\bf \hat{\mu}_{k}, \hat{\Sigma}_{k}})$ 
  in Equation \ref{eq:membership_weight} for each object.}}
\end{table*}

Our two methods estimate the parameters of the mixtures of 
Gaussian distributions differently. 
When using  method A to describe the color distribution, 
we use the Dirichlet process in a non-parametric Bayesian inference 
of the parameters for the mixtures as well as 
a Markov chain Monte Carlo (MCMC) algorithm \citep{Rasmussen99,Liverani15,Hastie15}.
Therefore, we acquire posterior distributions of the parameters 
such as $K$, $w_{k}$, $\mu_{k}$, and $\Sigma_{k}$ in  method A\footnote{We adopt a hyperparameter $\alpha ~=~ 2.2$ by checking the clustering 
results and their correspondence to the colors of the known taxonomy types 
\citep[see][for discussion]{Shin09}.}.
Method B adopts an expectation-maximization iterative algorithm 
to find the  parameters of the mixture models 
for a given number of mixtures \citep{Bishop06,Chen15}. 
We obtain the maximum-likelihood estimation of the parameters in  method B. 
When  method B updates the  assignment of mixture components 
over iterations in the expectation-maximization algorithm, 
the objects with known taxonomy types 
do not change their mixture assignments, 
and they work as an anchor in the color distribution 
for evaluating $w_{k}$, $\mu_{k}$, and $\Sigma_{k}$. 
We present the maximum a posteriori    (MAP) estimation of the parameters 
in method A below, and the best estimation of the parameters 
in method B corresponds to the maximum likelihood estimation.

We have to identify which taxonomic type matches each cluster 
in method A, whereas
identification of taxonomy types is consequently derived 
from a few objects with known taxonomic types in method B. 
Because we already know the colors of 318 objects 
with 12 known taxonomy types (A, B, C, D, K, 
L, Q, R, S, T, V, and X), 
we can determine the correspondence between the mixture 
components and the known taxonomic types 
in  method A.

\begin{table*}
\caption{Parameter estimation of mixture components 
in  method B\label{tab:methodB}}
\centering
\begin{tabular}{cccc}
\hline
\hline
Cluster (Taxonomy)\tablefootmark{a} & $w$ & $\mu$ & $\Sigma\tablefootmark{b}$ \\
\hline
1 (-) & 2.98E-02 & (2.49E-01, 4.09E-02, 7.46E-01) & 
$ \begin{pmatrix} 
1.17E-01 & -6.23E-02 & 2.42E-01 \\ 
\cdots & 7.04E-02 & -7.54E-02 \\ 
\cdots & \cdots & 6.16E-01 \\ 
\end{pmatrix} $\\
2 (-) & 1.02E-01 & (6.11E-02, 2.07E-02, 2.02E-01) & 
$ \begin{pmatrix} 
8.03E-03 & -1.91E-03 & 1.98E-02 \\ 
\cdots & 6.25E-03 & 2.20E-03 \\ 
\cdots & \cdots & 6.05E-02 \\ 
\end{pmatrix} $\\
3 (C) & 2.29E-01 & (5.94E-02, -2.25E-03, 1.63E-01) & 
$ \begin{pmatrix} 
4.35E-03 & 1.71E-04 & 1.13E-02 \\ 
\cdots & 1.97E-03 & 2.50E-03 \\ 
\cdots & \cdots & 3.24E-02 \\ 
\end{pmatrix} $\\
4 (D) & 6.97E-02 & (2.44E-01, 8.18E-02, 7.08E-01) & 
$ \begin{pmatrix} 
7.82E-03 & 2.05E-04 & 2.10E-02 \\ 
\cdots & 2.63E-03 & 3.31E-03 \\ 
\cdots & \cdots & 6.06E-02 \\ 
\end{pmatrix} $\\
5 (K) & 5.72E-02 & (2.67E-01, -4.73E-02, 6.98E-01) & 
$ \begin{pmatrix} 
2.31E-03 & -1.39E-04 & 5.60E-03 \\ 
\cdots & 5.99E-04 & 3.63E-04 \\ 
\cdots & \cdots & 1.52E-02 \\ 
\end{pmatrix} $\\
6 (L) & 7.87E-02 & (3.41E-01, -5.64E-03, 9.34E-01) & 
$ \begin{pmatrix} 
3.87E-03 & 5.37E-04 & 1.02E-02 \\ 
\cdots & 2.03E-03 & 2.92E-03 \\ 
\cdots & \cdots & 3.00E-02 \\ 
\end{pmatrix} $\\
7 (-) & 6.49E-03 & (2.73E-01, -1.88E-01, 5.39E-01) & 
$ \begin{pmatrix} 
2.50E-02 & -5.15E-04 & 6.07E-02 \\ 
\cdots & 4.45E-04 & -2.17E-03 \\ 
\cdots & \cdots & 1.51E-01 \\ 
\end{pmatrix} $\\
8 (-) & 3.77E-02 & (3.41E-01, -1.42E-01, 7.45E-01) & 
$ \begin{pmatrix} 
2.19E-03 & 4.46E-04 & 6.00E-03 \\ 
\cdots & 1.78E-03 & 2.41E-03 \\ 
\cdots & \cdots & 1.78E-02 \\ 
\end{pmatrix} $\\
9 (S) & 2.55E-01 & (3.22E-01, -9.15E-02, 7.85E-01) & 
$ \begin{pmatrix} 
3.69E-03 & 2.26E-04 & 1.02E-02 \\ 
\cdots & 1.48E-03 & 1.68E-03 \\ 
\cdots & \cdots & 2.98E-02 \\ 
\end{pmatrix} $\\
10 (-) & 2.64E-02 & (2.80E-01, 1.94E-02, 7.50E-01) & 
$ \begin{pmatrix} 
1.62E-03 & -1.12E-03 & 1.02E-03 \\ 
\cdots & 9.82E-03 & 1.02E-02 \\ 
\cdots & \cdots & 1.80E-02 \\ 
\end{pmatrix} $\\
11 (V) & 2.08E-02 & (3.61E-01, -2.04E-01, 7.39E-01) & 
$ \begin{pmatrix} 
8.97E-03 & 4.62E-03 & 2.70E-02 \\ 
\cdots & 2.11E-02 & 2.23E-02 \\ 
\cdots & \cdots & 8.68E-02 \\ 
\end{pmatrix} $\\
12 (X) & 8.84E-02 & (2.19E-01, -4.26E-03, 5.84E-01) & 
$ \begin{pmatrix} 
4.43E-03 & 2.28E-04 & 1.20E-02 \\ 
\cdots & 1.14E-03 & 2.17E-03 \\ 
\cdots & \cdots & 3.52E-02 \\ 
\end{pmatrix} $\\
\hline
\end{tabular}
\tablefoot{
\tablefoottext{a}{Types except for C, D, K, L, S, V, and X types 
are not shown here since their clusters do not seem physically meaningful.}
\tablefoottext{b}{Because $\Sigma$ is a symmetric matrix, we do not present 
a lower triangular part of matrix elements in this table.}}
\end{table*}

\begin{table*}
\caption{Taxonomy types in  method B \label{tab:methodB_membership}}
\centering
\begin{tabular}{ccccccc}
\hline
\hline
Object name &
Cluster number (Taxonomy type)\tablefootmark{a} & 
  \multicolumn{5}{c}{Membership weights\tablefootmark{b}} \\
& & (1) & 
(2) & $\cdots$ &
(11) & (12) \\
\hline
1999 NG53 & 9 (S) & 6.2E-04 & 3.3E-03 & $\cdots$ & 9.5E-05 & 4.4E-02 \\
Kenzo & 6 (L) & 2.5E-03 & 7.2E-04 & $\cdots$ & 2.9E-05 & 1.8E-01 \\
1999 NH54 & 8 (-) & 4.9E-03 & 2.2E-04 & $\cdots$ & 6.9E-02 & 2.7E-04 \\
2005 SA221 & 6 (L) & 1.2E-02 & 1.1E-04 & $\cdots$ & 4.0E-03 & 2.1E-03 \\
1999 RP116 & 10 (-) & 1.9E-01 & 1.6E-04 & $\cdots$ & 3.3E-08 & 3.9E-18 \\
3090 P-L & 4 (D) & 8.6E-03 & 2.9E-02 & $\cdots$ & 5.0E-06 & 6.7E-05 \\
1999 NO55 & 9 (S) & 7.1E-04 & 3.7E-03 & $\cdots$ & 5.5E-04 & 5.3E-04 \\
1999 XV242 & 9 (S) & 8.0E-04 & 3.2E-04 & $\cdots$ & 5.2E-04 & 7.4E-03 \\
Marlu & 3 (C) & 7.0E-04 & 1.1E-01 & $\cdots$ & 4.3E-11 & 2.3E-02 \\
Sansyu-Asuke & 9 (S) & 5.2E-04 & 1.3E-03 & $\cdots$ & 4.4E-04 & 6.0E-04 \\
1996 VO4 & 2 (-) & 1.4E-02 & 5.4E-01 & $\cdots$ & 5.3E-04 & 2.0E-03 \\
1999 VJ14 & 9 (S) & 1.3E-02 & 4.2E-03 & $\cdots$ & 5.0E-04 & 2.5E-11 \\
Priestley & 9 (S) & 1.4E-02 & 3.1E-05 & $\cdots$ & 8.8E-03 & 2.7E-04 \\
Halaesus & 4 (D) & 4.0E-03 & 3.4E-03 & $\cdots$ & 7.3E-07 & 7.0E-04 \\
Belinskij & 3 (C) & 7.1E-03 & 2.8E-01 & $\cdots$ & 4.8E-04 & 2.1E-01 \\
2000 HC36 & 9 (S) & 1.3E-03 & 5.8E-04 & $\cdots$ & 2.4E-03 & 2.3E-02 \\
2000 HA41 & 4 (D) & 4.2E-03 & 7.6E-03 & $\cdots$ & 2.0E-07 & 3.0E-01 \\
1992 SU & 3 (C) & 1.4E-03 & 1.4E-01 & $\cdots$ & 3.1E-06 & 8.7E-02 \\
Neuvo & 9 (S) & 1.4E-03 & 1.7E-04 & $\cdots$ & 2.8E-03 & 2.7E-03 \\
2000 GT136 & 6 (L) & 1.5E-02 & 7.7E-03 & $\cdots$ & 1.2E-05 & 4.0E-07 \\
\hline
\end{tabular}
\tablefoot{This table is published 
in its entirety in  machine-readable format.
A portion is shown here for guidance regarding its form and content.
\tablefoottext{a}{Types except for C, D, K, L, S, V, and X types 
are not shown here  since their assignments are not reliable in the clustering results.}
\tablefoottext{b}{The membership weights are defined by 
  $\hat{w}_{k}  N({\bf x} | {\bf \hat{\mu}_{k}, \hat{\Sigma}_{k}})$ 
  in Equation \ref{eq:membership_weight} for each object.}}
\end{table*}

\subsection{Results}

The taxonomy assignment (i.e., cluster membership) 
in method A depends on a posterior dissimilarity 
matrix of mixture assignments derived from the MCMC samples.
The optimal determination of cluster membership corresponds to the case of 
partitioning around medoids on the dissimilarity matrix \citep{Liverani15}. 
We find six clusters in this cluster membership assignment, and we interpret 
four of them as C, S, V, and X types because of their color distributions 
and their similarity to those of known taxonomy samples  shown 
in Figure \ref{fig:methodA_membership}. Table \ref{tab:methodA_raw_membership} 
shows the results of this taxonomy assignment (hereafter method A1).

We also produce a different set of cluster 
memberships and taxonomy types incorporating the MAP 
estimation of the mixture parameters in  method A.
We present the mixture parameters 
$w_{k}$, $\mu_{k}$, and $\Sigma_{k}$ for the resulting 13 
mixture components in the MAP estimation of  method A 
(see Table \ref{tab:methodA_MAP}). 
This cluster 
membership assignment given in Table \ref{tab:methodA_MAP_membership} is determined by 
finding the maximum posterior of component inclusion 
(i.e., membership weight) 
\begin{equation}\label{eq:membership_weight}
\underset{k}{\arg\max} ~ 
  \hat{w}_{k}  N({\bf x} | {\bf \hat{\mu}_{k}, \hat{\Sigma}_{k}}),
\end{equation}
for each data with the estimated parameters 
$\hat{w}_{k}$, $\hat{\mu}_{k}$, and $\hat{\Sigma}_{k}$. 
This method 
results in 13 mixtures where only four 
clusters correspond to the meaningful taxonomy types C, S, V, and X 
(hereafter, method A2; see Table \ref{tab:methodA_MAP_membership}). 
The cluster membership in the MAP estimation is different 
from that derived from the dissimilarity matrix.
While the MAP estimation simply works as a point estimation 
of the parameter values, the cluster membership assignment 
using the dissimilarity matrix embraces entire information 
in the posterior MCMC samples.

The numbers of 
consistent assignments between the two ways in  method A 
are 1551, 1774, 17, and 260 for C, S, V, and X types, respectively. 
These numbers correspond to approximately 37\%, 42\%, 0.40\%, and 
6.2\% for C, S, V, and X types, respectively, or 3602 of the total 
4213 objects.
We expect these objects with the consistent assignments to have 
more reliable taxonomy assignments than others.

Method B offers different taxonomic types with its 
estimation of the maximum posterior of component inclusion 
(i.e., membership weight). 
Although we consider the 12 known taxonomic types 
by including their known samples, the mixtures found in  method B 
appear to have meaningful structures corresponding to only the 
seven taxonomy types (C, D, K, L, S, V, and X)   shown 
in Figure \ref{fig:methodB_membership} and   summarized 
in Table \ref{tab:methodB}.

Table \ref{tab:methodB_membership} presents the taxonomic types 
found in  method B with the normalized membership weights 
for the estimated 12 mixtures even though the only 7 mixtures 
display valid structures in the color space. If a single cluster 
membership weight dominates the others, the type assignment is quite 
reliable. For example,  object 1999 NO55 has the largest and dominant 
cluster membership weight (0.93) for   cluster 9, which corresponds to 
 taxonomy type S (see Table \ref{tab:methodB_membership}). 
The taxonomic classification of  object 2000 HA41 is not strongly 
supported by  method B since its largest cluster membership weight 
is merely 0.31 for  taxonomic type D. Among the 4213 objects used 
as input data, the number of objects with  weights higher than 
0.9 in the membership assignment by  method B are 30, 29, 0, 23, 306, 17, 
and 0 for  types C, D, K, L, S, V, and X, respectively. When counting 
the objects with a membership assignment mixture weight higher than 
0.5, we recover 1093, 252, 129, 240, 1022, 57, and 235 objects for the 
types C, D, K, L, S, V, and X, respectively.

The mixture parameters derived here (Tables \ref{tab:methodA_MAP} and 
\ref{tab:methodB}) can be used to infer taxonomic types of objects 
for newly acquired color measurements. 
For example, if we suppose that for  object 1999 NG53 we acquire
a new measurement of color (0.28, -0.06,  0.69), 
which is actually what we include as the input data 
for clustering, the new measurement becomes 
a new ${\bf x}$ in Equation \ref{eq:membership_weight}, 
and we can estimate the membership weights of taxonomic types 
for the new color with the derived parameters $\hat{w}_{k}$, $\hat{\mu}_{k}$, 
and $\hat{\Sigma}_{k}$ in either Table \ref{tab:methodA_MAP} 
or \ref{tab:methodB}. 
If the new color measurement is in a sub-dimension such as 
$g - i$ or $i - z$, we can still use the derived parameters 
to infer taxonomic types in terms of a marginalized distribution 
of the estimated Gaussian mixture distributions.

\begin{table}
\caption{Objects with the consistent taxonomy types in  methods A and B}
\label{tab:method_ensemble_membership}
\centering
\begin{tabular}{c c}
\hline
\hline
Object name & Taxonomy type\\
\hline
1999 NO55 & S \\
1999 XV242 & S \\
Marlu & C \\
Sansyu-Asuke & S \\
1999 VJ14 & S \\
Belinskij & C \\
2000 HC36 & S \\
1992 SU & C \\
Neuvo & S \\
2000 HE68 & C \\
2000 HK80 & S \\
2000 HU37 & C \\
2001 TD154 & C \\
2000 GK142 & S \\
2001 PZ6 & S \\
2000 GB164 & C \\
2000 HR80 & S \\
Begonia & C \\
2001 QV268 & S \\
\hline
\end{tabular}
\tablefoot{Table \ref{tab:method_ensemble_membership} is published 
in its entirety   in  machine-readable format.
A portion is shown here for guidance regarding its form and content.}
\end{table}

Combining the three clustering membership results given in Tables 
\ref{tab:methodA_raw_membership}, \ref{tab:methodA_MAP_membership}, and \ref{tab:methodB_membership} 
helps us sort out the most reliable taxonomic type assignments for 
the C, S, V, and X types  considered in the three different results. 
One simple way of 
ensemble learning \citep{ensemble_intro} is 
accepting only the consistent taxonomic assignments among the three 
different assignments inferred by methods A and B
\citep[see][for a review]{ensemble,ensemble_clustering_ext,ensemble_clustering}. In this way we can 
select the objects with the most reliably inferred taxonomic types 
\citep[e.g.,][]{Shin18}. Table \ref{tab:method_ensemble_membership} presents 
the objects with the C, S, V, and X taxonomic types that are in agreement among 
the three different inferences. 

Among the 4213 objects we find 1176 (28\%), 
1104 (26\%), 16 (0.38\%), and 0 (0\%) objects with 
the consistent taxonomic assignments for 
C, S, V, and X types, respectively. Focusing on objects with X type 
in  the method A assignment, which means those that appear consistently 
in X-type taxonomy of  method A1 and A2, D and L types correspond to 180 
and 60 objects in  method B assignment, respectively 
(see Figure \ref{fig:consistent_plot}). K- and X-type objects in  method B 
do not belong to the group of X-type objects inferred in the method A 
assignment. Since  method B can cluster minor types such as D, L, and K 
separately from the large X-type cluster while method A cannot, there is not 
always agreement between methods A and B for objects assigned as X type 
in method A.

The lack of consistent X-type assignments between  methods A and B 
is not a surprising result when we check taxonomy inference of the 318 
Bus-DeMeo samples in  method A. These samples are not used in the training 
of  method A because  method A is an unsupervised learning method. 
For 44, 173, 13, and 31 samples of C, S, V, and X types, 
respectively, in the Bus-DeMeo data we use  method A2 to derive 
the taxonomy assignments with the derived MAP mixture parameters (i.e., 
Table \ref{tab:methodA_MAP}). The derived 
taxonomy types of 42, 172, 13, and 0 objects are the same as the 
Bus-DeMeo taxonomy assignment for C, S, V, and X types, respectively. 
The color ranges of the mixture associated with X type is found 
to be strongly intruded by other C, S, and V types in  method A 
(see Figures \ref{fig:methodA_membership} 
and \ref{fig:methodB_membership}). 
Checking consensus between  methods A and B affirms that the X-type 
assignment in  method A is not conclusive while C-, S-, and V-type 
assignments in  method A is more reliable than the X-type assignment.

Since our method A belongs to the unsupervised learning, our result 
(in particular method A2) cannot be shown as a supervised learning 
method \citep[e.g.,][]{Carvano10,Erasmus18,Erasmus19};  
the result has  been tested and presented 
with the assumption that the samples used in training are complete
and distributed in the same way as the unknown test samples \citep[see][for discussion]{Dundar07}. 
Instead, we 
can present the fraction of the Bus-DeMeo reference objects, which are 
not included in the training step of  method A2, for the correct and 
incorrect clustering assignment in terms of the known taxonomy. 
As estimated from the numbers mentioned above, 
the fraction of correct clustering assignments, 
which is similar to classification accuracy in supervised learning, 
is about 95.5, 98.9, 100.0, and 0\% for C, S, V, and X types, 
respectively, in  method A2. 

We also inspect how  method A2 assigns clustering membership 
for the Bus-DeMeo samples in taxonomy groups 
not represented by  method A2 (i.e., A, B, D, K, L, Q, R, and T). 
All B-type samples appear to be members of the C-type cluster. For 
the D-type samples, the dominant 9 objects among 13 samples are assigned 
to the X type which is mainly influenced by the C, S, and V clusters 
in  method A2 
(see Figures \ref{fig:methodA_membership} and \ref{fig:methodB_membership}).
The single R-type object in the Bus-DeMeo samples is 
included in the V cluster by  method A2.  The S cluster includes 
the most objects of the other types. Therefore, the precision 
estimation of the correct taxonomy assignment for the C, S, V, and X become 
63.6, 77.1, 92.9, and 0\%, respectively. We note that these numbers 
cannot be interpreted as precision presented for supervised learning results.

\section{Discussion and conclusion}

\subsection{New features}

\begin{figure*}
  \centering
  \includegraphics[width=\hsize]{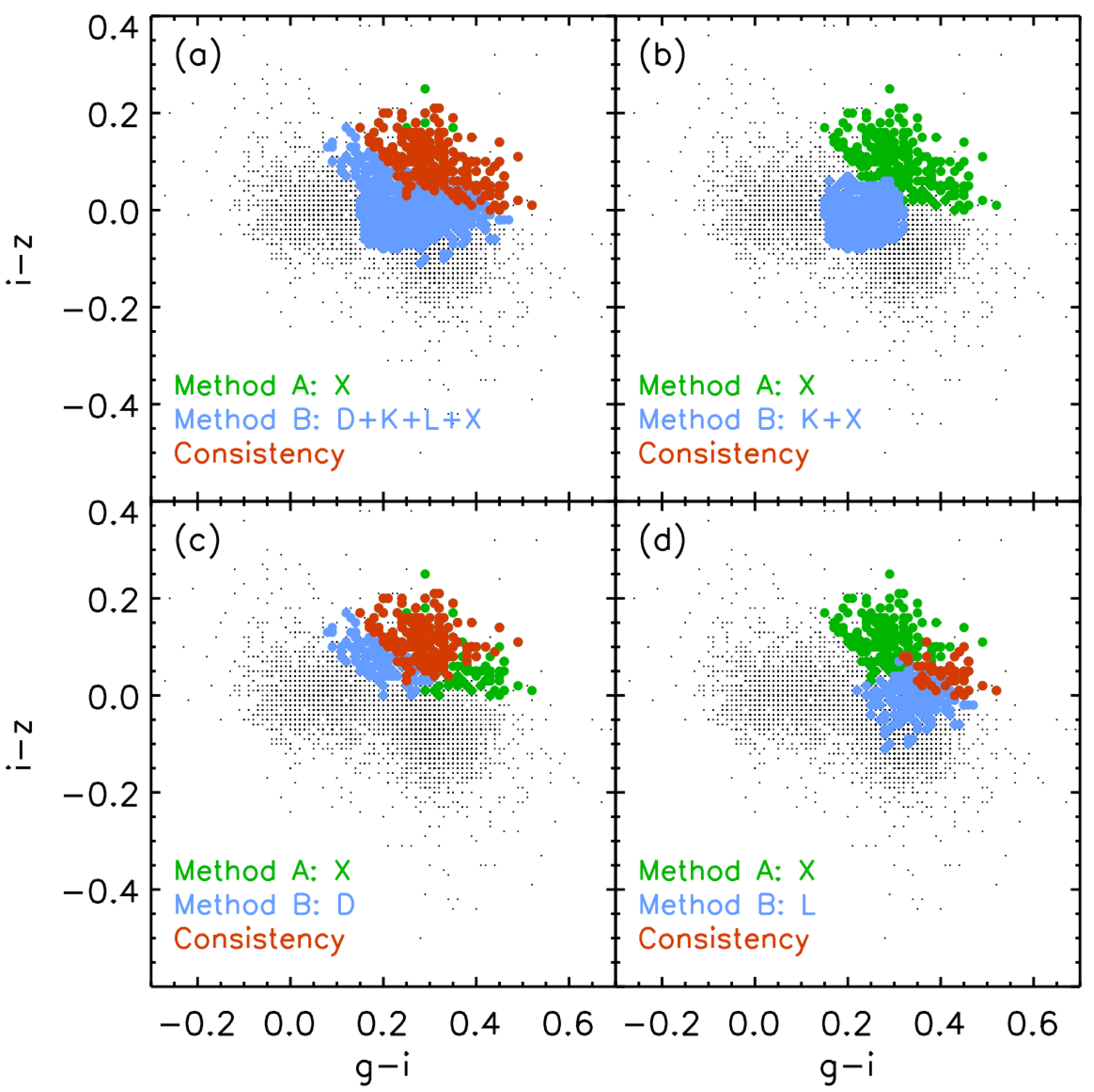}
  \caption{Two-dimensional projection plots for consistent taxonomy objects. Different colors correspond to X-type assignments in  method A (green; consistent objects in method A1 and A2), comparative taxonomy types in  method B (blue), consistent objects between  method A and B assignments (red), and all our samples (black). D- and L-type objects in method B correspond to a large portion of objects with X type in the method A result, whereas K and X types in method B do not correspond to X type in method A.}
  \label{fig:consistent_plot}
\end{figure*}

\begin{figure*}
  \centering
  \includegraphics[width=0.9\hsize]{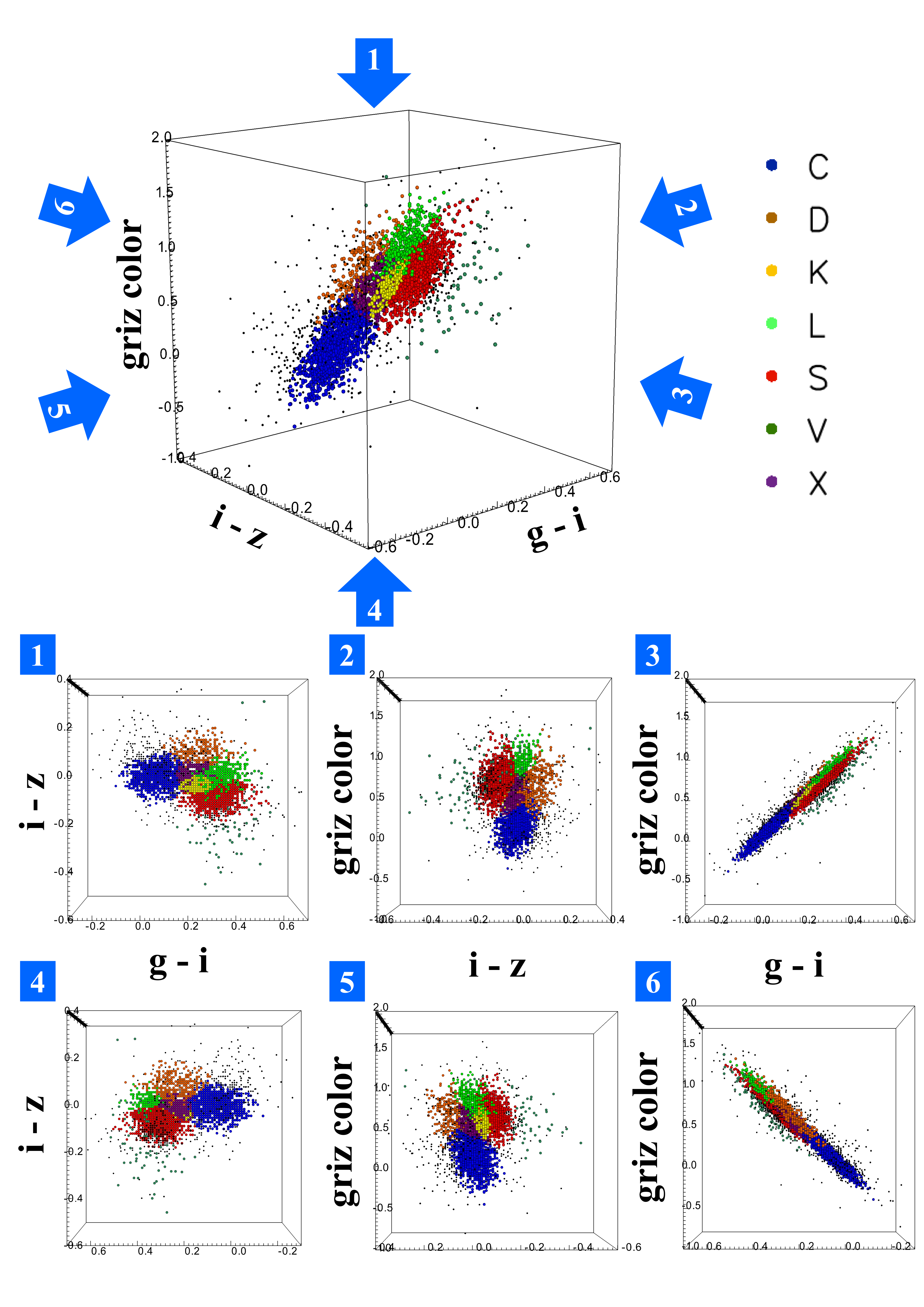}
  \caption{Perspective view of the 3D clustering results in method B. The  color-coding is the same as in Fig. \ref{fig:methodB_membership}. Different colors correspond to different taxonomy classification memberships for C (blue), D (orange red), K (yellow), L (lime), S (red), V (lime green), X (purple), and Unassigned (black).
The overlapping members of each type in the 2D plane seem to be separated by relatively clear boundaries in 3D space.}
  \label{fig:3D_perspective_view}
\end{figure*}

In previous studies taxonomic classification of asteroids in SDSS was done in 2D parameter space using the slope (or g-i) and the absorption depth (or i-z), for instance. 
In this paper we introduced an additional parameter, the griz color, which is the flux value of the normalized reflectance in the SDSS bands. 
As is evident in Figures \ref{fig:methodA_membership}, \ref{fig:methodB_membership}, \ref{fig:consistent_plot}, and \ref{fig:3D_perspective_view}, 
the griz color is not orthogonal to (g-i) and (i-z) in the newly defined color space. Consequently, as the slope (or g-i) grows, the area that it makes becomes larger; in the meantime, as the absorption depth (or i-z) increases it becomes smaller. 
This color exhibits a significantly wide distribution ranging from –1.0 to +2.0.
As shown in Figure \ref{fig:3D_perspective_view}, the distribution of 
taxonomies is different depending on which direction we look 
in 3D space.
We note that in Figure \ref{fig:3D_perspective_view} the   X and K types are both visible in the middle from the +griz direction, 
while only X type is seen from the –griz direction as K type is hidden from view. The same applies for -(g-i) and +(g-i), in parallel, -(i-z) and +(i-z) directions.

Likewise, most of the  surface area of  L type is clearly visible in the +griz direction; on the other hand, a fairly large fraction of L-cloud is obscured by S-, D-, and X-clouds. Because of the overlapping nature of the spatial distribution, we cannot completely eliminate a level of uncertainty in asteroid taxonomy, for the present. If we have a sufficient number of spectra to be used for reference points, we should be able to make a clearer division. Interactive plots are made available on the website \footnote{\label{web}\url{https://data.kasi.re.kr/vo/asteroid_taxonomy/}} where the 3D structure of the clouds can be explored.
As (g-i), (i-z), and the griz colors display fairly continuous distribution, the 3D structure (the cloud of data points) that the three colors create also exhibits this property. 
It is naturally due to the continuously changing nature of the reflectance spectra. 
Remarkably, the 2D swarms revealed in various forms of principal component diagrams  has proven to be the ``shadow'' of the 3D structure projected on the 2D ``floor.''

\subsection{Objects without assigned taxonomic types: Their true nature}
The black dots that are lying outside the color-coded (taxonomy-assigned) clusters in the relevant figures are the unassigned data points. 
Such objects without assigned taxonomic types (hereafter OWATs) are currently unidentifiable, and we do not know their physical nature; nevertheless, the newly adopted machine learning method is based on Bus-DeMeo taxonomy system.
In order to check if the distribution of such OWATs have relevance to photometric error, we plotted the data according to error, yet we did not find any clear and systematic trend. Hence, it would be reasonable to conclude that they indeed exist, and that we need to explore and understand their true physical nature in this 3D color space.

In order to discover their true nature, we should  obtain reflectance spectra of those OWATs and  match their spectra with meteorite analogs for precise identifications. 
We expect the 3D taxonomy to evolve as we assign taxonomy to and discover the nature of these OWATs in the taxonomically  and geophysically unexplored (hence unidentified) territories in the 3D color space. 

\subsection{Models and training data}
We suggest multiple ways to use our clustering results with the inferred 
taxonomy types. First, if people want to pin down the most 
reliable C-, S-, V-, and X-type objects, we recommend   using 
Table \ref{tab:method_ensemble_membership} to identify them. Second, 
when people want to identify various taxonomic types (in particular 
C, D, K, L, S, V, and X types) or choose objects 
with a certain reliability threshold, they need to use  Table 
\ref{tab:methodB_membership}. Third, objects with 
highly uncertain taxonomy assignments might be interesting targets for 
further studies, and recognizing them requires either an inspection of the 
disparity between Tables \ref{tab:methodA_raw_membership} and 
\ref{tab:methodB_membership} or an examination of the low-probability 
taxonomy assignments presented in Table \ref{tab:methodB_membership}.

Taxonomic types that overlap in 2D space 
appear to be able to be distinguished in 3D space. 
This is a better reflection of the parametric representation 
of each taxonomy and shows that it is useful 
for determining taxonomic classification of asteroids. 
However, the accuracy of the taxonomy determined by this method 
is not guaranteed. It is necessary to confirm the spectroscopic 
observation results. 
Nevertheless, at the present time it is significant that the results of 
statistical approaches using the finite spectral samples can probably 
determine taxonomies of asteroids for their photometric results. 

\begin{figure*}
  \centering
  \includegraphics[width=0.9\hsize]{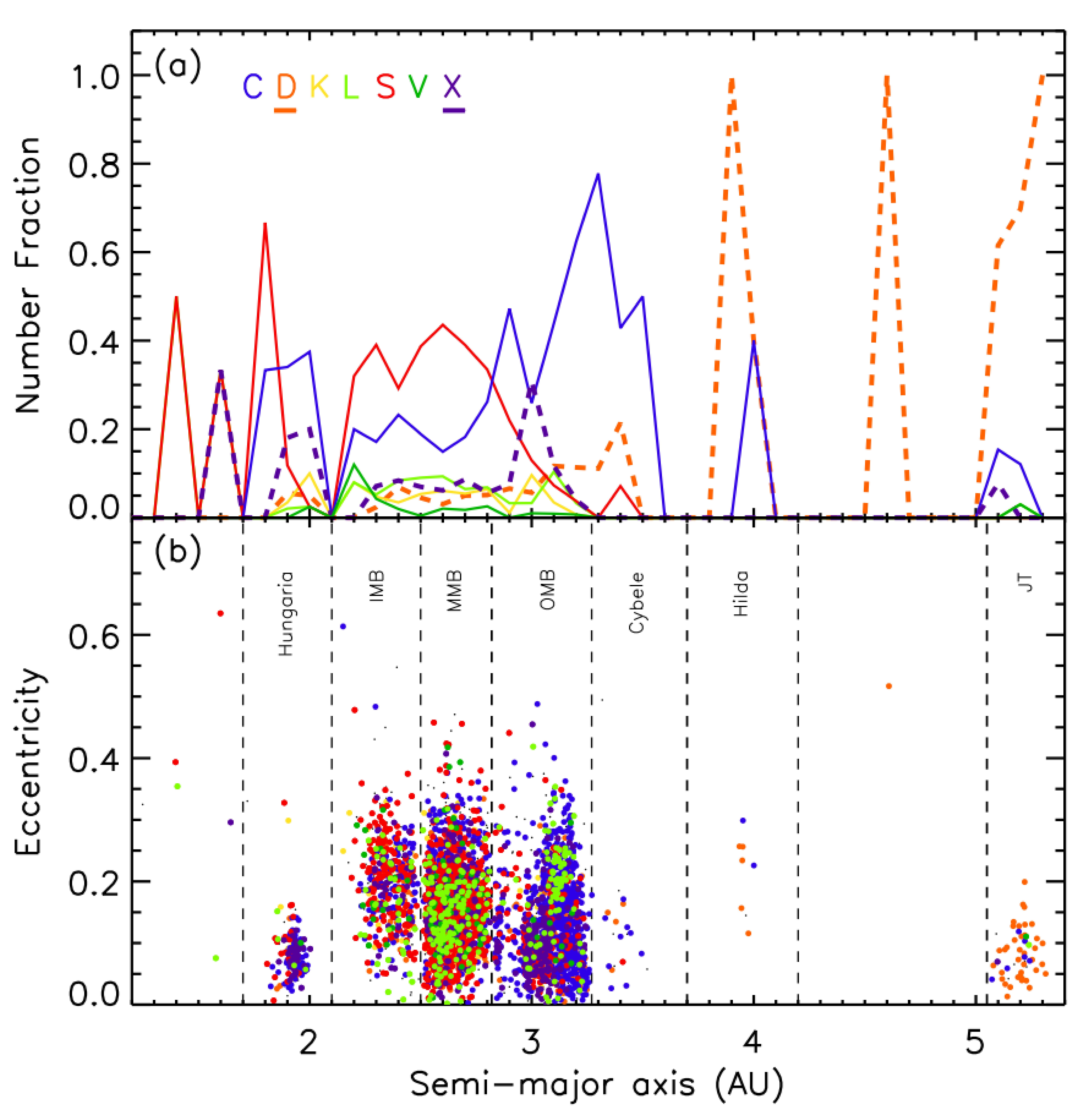}
  \caption{Spatial distribution of 4213 SDSS MOC4 asteriods. (a) Biased fraction of each type in each 0.1 AU bin. (b) Taxonomy distribution of asteroids in proper orbital elements plane (semi-major axis vs. orbital eccentricity).}
  \label{fig:Fraction}
\end{figure*}

Increasing the size of the photometric training samples plays an 
important role in improving the clustering results of method A 
where the number of training samples and the density of the the samples 
in the color space affect the quality of the clustering 
results. When we collect more photometric samples, we expect to unveil 
diverse structures 
in the color space including peculiar or unknown taxonomy populations, 
and to define color boundaries with certainty  for major populations such as 
C, S, V, and X types. Future surveys such as the Legacy Survey of Space and Time 
(LSST) will produce multi-band optical data for 
the unprecedented large number of asteroids that can cover the diverse 
color populations \citep{LSST}.

Gathering more spectroscopically confirmed samples will also play a critical role 
in improving the results from our clustering methods. 
In particular, the added data with confirmed taxonomic types 
will substantially improve the
reliability of the results from method B, 
which explicitly uses data with known taxonomic types in clustering 
\citep[see][for discussion]{Cozman03}. 
We cannot find indicative structures of some taxonomic types (e.g.,  
A, B, Q, R, and T) in method B due to an insufficient sample size 
of objects with already known types. The number of  known taxonomy samples for these types is less than six, and method B fails to infer the cluster structure with such a small number of samples. The result of method 
B also demonstrates that the membership determination for the types 
K and X is not as confident as for other types (e.g.,  C, D, L, S, 
and V). Future spectroscopic observations such as the proposed mission CASTAway 
\citep{Bowles18} can substantially increase the number of spectroscopically confirmed taxonomy samples, covering a broad range of taxonomy types.

We expect others to use the published information on the trained models such as the cluster membership probabilities and model parameters in methods A and B with their own prior probability or model results. In particular, 
people may conceive of a new way to combine the results of  models A and B instead 
of simply checking for agreement between the two models \citep[e.g.,][]{NGUYEN2020107104}. 
For example, a stacking approach \citep{Wolpert92} can be adopted to estimate a better 
method of combining our multiple results or combining our results with other taxonomy assignment results. The Bayesian estimation of 
a new posterior probability for taxonomy assignment is also possible in combining 
our model likelihood with other prior probabilities or in updating our model 
likelihoods. We provide simple Python scripts that can be used to infer 
asteroid taxonomy for given colors,  our taxonomy result tables,  as well as 
the interactive plots at our website.

\subsection{Implications for taxonomic distribution of asteroids}
We plot the spatial distribution of 4213 asteroids with assigned taxonomy from method B in Figure \ref{fig:Fraction}. The semi-major axis bins we chose are 0.1 AU wide ranging from 1.2 to 5.4 AU from the sun. The MOC4 objects in Hungarias (1.78~-~2.05 AU); the inner (2.05~-~2.5 AU), the middle (2.5~-~2.82 AU), and the outer (2.82~-~3.27 AU) mainbelt (IMB, MMB, and OMB, respectively); Cybeles (3.27~-~3.7 AU); Hildas (3.7~-~4.2 AU); and Jupiter Trojans (JT, 5.05~-~5.40 AU) are included in this study. 
We then calculated the biased fraction of asteroids in each bin in Figure \ref{fig:Fraction} (a), and plot the objects in semi-major axis versus eccentricity proper orbital element plane in Figure \ref{fig:Fraction} (b). 
We note that two-thirds of the asteroids with low orbital inclinations are excluded in Figure \ref{fig:Fraction} as we applied the galactic latitude cutoff. We classified six objects outside these regions, five in the near-Earth object (NEO) and Mars Crosser (MC) regions, and 1 near $\sim$ 4.6 AU between the Hilda and Trojan regions. For a larger sample of NEOs and MCs in SDSS see \citet{Carry16}.

In these figures the distribution of asteroids in Hungarias and IMB-MMB regions is dominated by S type (11 \%, 33 \% and 41 \% in Hungarias, IMB, and MMB, respectively) out to 3 AU, while C type takes the position in OMB (45.6 \%) and beyond. The apparent lack of X-type asteroids in the Hungaria family is due to the exclusion of low orbital inclination objects as we applied a more stringent galactic latitude cutoff. 
While X-type asteroids occupy a large fraction of Hungarias (17.7 \%), the fraction drops in the IMB (7.3 \%)~ and ~MMB (6.8 \%) sections to suddenly increase ($\sim$ 30 \%) in the inner OMB. X type is known to be composed of three subtypes, E, M, and P \citep{Tholen84}, where P types show the lowest albedo with a featureless reddish spectra. 
They are found mostly in OMB and beyond \citep{Lazzarin95} with an apparent peak at 4 AU \citep{McSween99}. 
Nevertheless, we find the peaks of X and P types to be located at 3~-~3.2 AU in Figure \ref{fig:Fraction} of this work and Figure 11 of \citet{DeMeo13}, although a one-to-one comparison is difficult as we do not separate E, M, and P types in this study.

At the same time, the V-type fraction grows in the IMB where the Vesta family dominates, while it becomes almost negligible in OMB. 
We then calculated the observed fraction of some minor taxonomy types such as D, K, and L across the main belt. 
The portion of D type in the outer OMB is between 10 \% and 20 \% (e.g., from one-fourth to one-eighth of the C-type fraction in this section); the fraction dramatically increases among Hilda (62.5 \%) and peaks at JT swarms (72.2 \%) to become an absolute majority. 
However, it is significant that D-type asteroids are also discovered in the innermost asteroid zone such as Hungarias (5.4 \%) and IMB (5.1 \%) \citep[see][]{DeMeo13,DeMeo14}. 
On the other hand, both K and L types are relatively evenly distributed in Hungarias and across the mainbelt, even if their contribution is not very significant.
The observed L-type fraction is $\sim$ 10 \% throughout the 2.2~-~2.7 AU region, while the K-type fraction demonstrates two less prominent peaks at 2.0 and 3.0 AU in our binning scheme. Due to the relatively small number of the sample (4123), hence sparsely populated data points in the orbital parameter plane, study of the dynamical structure of the asteroid belt is rather difficult. If we expand the sample with either better photometric qualities or spectroscopic measurements, a higher resolution picture of the dynamical families should be revealed \citep{Ivezic02, Parker08}.

Our results are generally consistent with those found in \citet{DeMeo13} for the biased results shown in their Figure 10 and with other earlier works \citep{Bus02b, Mothe03}. For example, S-types dominate the IMB and MMB by number, and the switch to C types being more populous occurs in the OMB. Discrepancies between our work and previous studies are attributable to  more low-inclination objects being excluded from this work due to our more stringent data quality cutoff including galactic latitude constraints and to the small sample sizes among Hildas and Jupiter Trojans. These discrepancies include a smaller contribution of V types (Vesta family) in the IMB and K types (Eos family) in the OMB, and different relative fractions of X and D in the Hildas and Jupiter Trojans. A larger sample  size for Hildas and Trojans than is available for SDSS has been studied by NEOWISE \citep{Grav12a,Grav12b}.

\subsection{The future of asteroid taxonomy}
In this work, we put forth a method for the taxonomic classification 
of asteroids based on the clustering analysis of the photometric data. 
This classification scheme can also be simply represented by a triplet 
or multiplet of photometric colors, either in LSST 
or in Johnson-Cousins photometric systems. Applying 
our methods with observation data acquired in these 
bands may allow taxonomic identification of  
interesting populations. Colors in these different bands may help us identify minor 
taxonomy groups that are not strongly concentrated in 
the SDSS dataset.

Including NIR colors in clustering analysis will be one 
way of extending our methodology to cover NIR taxonomy 
classification. \citet{Popescu18} presented possible taxonomy classification 
in the $J - K_{s}$ and $Y - J$ color-color space by adopting  straight line 
boundaries among the types. We plan to investigate the optical-NIR combined 
clustering analysis of the objects studied in this paper, generating 
the Gaussian mixture models in color space over the optical--NIR colors and 
comparing the results from the current analysis and the combined analysis.

Our taxonomy method is extensible for many asteroid studies. 
We can combine the study of space weathering trends in S-complex subtypes 
with their color distribution of 3D parameter space. 
Furthermore, we are cautiously optimistic that the taxonomic distributions 
of asteroid families in proper orbital element space 
may reveal a more detailed interpretation of their origin and evolution.

\begin{acknowledgements}
This study is supported by Korea Astronomy and Space Science Institute. FED acknowledges funding from the National Aeronautics and Space Administration under Grant nos. 80NSSC18K0849 and 80NSSC18K1004 issued through the Planetary Astronomy Program. We would like to thank anonymous reviewer for his/her careful reading of our manuscript and insightful comments with valuable suggestions.
\end{acknowledgements}

\bibliographystyle{aa} 
\bibliography{ntaxref} 

\end{document}